\documentclass[12pt]{article}
\usepackage{graphicx,amsmath,amssymb,color,bm}

\begin{document}

\centerline {\Large\textbf {The effect of perpendicular electric field on }}
\centerline {\Large\textbf {Temperature-induced plasmon excitations for intrinsic silicene}}


\centerline{Jhao-Ying Wu$^{1,\star}$, Chiun-Yan Lin$^{1}$, Godfrey Gumbs$^{2,\dag}$, and Ming-Fa Lin$^{1,\dag\dag}$ }
\centerline{$^{1}$Department of Physics, National Cheng Kung University,
Tainan, Taiwan 701}
\centerline{$^{2}$Department of Physics and Astronomy, Hunter College at the City University of New York,\\ \small }
\centerline{695 Park Avenue, New York, New York 10065, USA}

\vskip0.6 truecm

\noindent

We use the tight-binding model and the random-phase approximation to investigate the intrinsic plasmon in silicene. At finite temperatures, an undamped plasmon is generated from the interplay between the intraband and the interband-gap transitions. The extent of the plasmon existence range in terms of momentum and temperature, which is dependent on the size of single-particle-excitation gap, is further tuned by applying a perpendicular electric field. The plasmon becomes damped in the interband-excitation region. A low damped zone is created by the field-induced spin split. The field-dependent plasmon spectrum shows a strong tunability in plasmon intensity and spectral bandwidth. This could make silicene a very suitable candidate for plasmonic applications.

\vskip0.6 truecm

$\mathit{PACS}$: 73.22.Lp, 73.22.Pr

\newpage

\bigskip

\centerline {\textbf {I. INTRODUCTION}}%

\bigskip

\bigskip

A carbon nanosheet, namely  graphene, was discovered a few years ago. The unique low-lying linear energy dispersions have provided researchers with an abundance of new physics, including the Klein paradox for tunneling through a rectangular electrostatic potential barrier \cite{Katsnelson:2006,Bai:2007,Young:2009}, the anomalous
quantum Hall effect \cite{Zhang:2005,McCann:2006}, and peculiar optical properties \cite{Abergel:2007,Nair:2008,Bonaccorso:2010,HoYH:2010,Zheng:2012}. Following the successes achieved with graphene, researchers have intensified their search for other graphene-like two-dimensional materials \cite{Wang:2012,Xu:2013,Butler:2013}. In recent years, silicene, a nanosheet consisting of silicon atoms, has been synthesized \cite{Vogt:2012,Chen:2012,Liu:2014}. Belonging to the same group as graphene in the periodic table, silicene is predicted to exhibit the feature-rich
electronic properties. Furthermore, it could be better suitable for practical applications than graphene for two main reasons. The first is its compatibility with current Si-based device technologies. The second is its sizable energy gap, owing to the larger magnitude of its atomic spin-orbit interaction
(SOI) and its buckled structure. The buckling means that the two sublattices are separated by a vertical distance of $2l$ ($\approx0.46{\AA}$), which is a result of the larger ionic size of silicon atoms compared to carbon atoms. Other group-IV crystals, e.g., Ge and Sn, are also expected to appear in this form in their 2D structures \cite{Takeda:1994,CChen:2012}.

The out-of-plane buckling suggests that an on-site potential difference between the $A$ and $B$ sublattices is tunable under a perpendicular electric field \cite{NiZ:2012,Lian:2015}. This allows for the control of the band gap, along with the spin and valley polarized states. Accordingly, many spin and valleytronics, like the quantum spin Hall effect \cite{Tabert:2013,Liu:2011,Tahir:2013} and the optically-generated
spin-valley-polarized charge carriers \cite{Ezawa:2012,Stille:2012,TTabert:2013,TTTabert:2013}, may be investigated. Additionally, when the electric potential exceeds the magnitude of the SOI, 2D silicene is predicted to undergo a topological phase transition from a topological insulator to a band insulator, a property that should be important in electronics and optics.

Collective-Coulomb excitations, dominated by the electron-electron interactions, are useful for understanding the behavior of electrons in a material due to the involvement of screening effects. This basic property has been widely studied in various low-dimensional systems, e.g., carbon nanotubes \cite{Shung1994,Chuu1997,Dmitrovi2008}, graphene layers \cite{Wu:2011,Lin:2012,Chuang:2013,Wu:2014,Luo2013,Javier2014,Grigorenko2012}, and other novel materials \cite{Scholz:2013,Stauber:2014,Pietro2013}. In monolayer graphene, plasmons (quantized collective-excitation modes) hardly exist at low frequencies due to the lack of free carriers. This may be improved by doping or insertion of a gate to increase the free-charge density \cite{Wunsch2006,SH2009}, i.e., changing the Fermi level in an extrinsic condition. Alternatively, an intrinsic low-frequency plasmon (with zero Fermi energy) may be induced by increasing the thermally excited electrons and holes in the conduction and valence bands, respectively \cite{Lin:2000}. This is based on the strong dependence of the carrier density on the temperature ($n\propto T^{2}$). The temperature-induced plasmon is also predicted to exist in monolayer silicene \cite{WuNJ:2014}. The difference is that the temperature-induced plasmon in graphene is always located in the interband region and suffers the Landau damping \cite{Lin:2000}. However, in silicene, the temperature-induced plasmon could be in a single-particle-excitation (SPE) gap arising from the spin-orbit interaction \cite{WuNJ:2014}. The opening of the SPE gap makes the plasmon undamped. In this paper, we investigate the effects of a perpendicular electric field on the T-induced plasmon in silicene. The energy-loss function is used to derive the plasmon spectrum. Three plasmon modes with distinct degrees of Landau damping are found in different momentum, temperature, and field strength ranges. The occurrences of plasmons and their discontinuous dispersions are associated with the special structures in the dielectric functions.

\bigskip
\bigskip
\centerline {\textbf {II. METHODS}}%
\bigskip
\bigskip

Similar to graphene, silicene consists of a honeycomb lattice of silicon atoms with two sublattices made up of A and B sites. The difference is that silicene has a buckled structure, with the two sublattice planes separated by a distance of $2l$ with $l=0.23$ ${\AA}$ (Fig. 1). In the tight-binding approximation, the Hamiltonian for silicene in the presence of SOI is written as \cite{CChen:2011,2Ezawa:2012}:
\begin{equation}
H=-t\sum_{\langle ij\rangle\alpha}c^{\dag}_{i\alpha}c_{j\alpha}+i\frac{\lambda_{SO}}{3\sqrt{3}}\sum_{\langle\langle ij\rangle\rangle\alpha\beta}v_{ij}c^{\dag}_{i\alpha}\sigma^{z}_{\alpha\beta}c_{j\beta}-i\frac{2}{3}\lambda_{R2}\sum_{\langle\langle ij\rangle\rangle\alpha\beta}u_{ij}c^{\dag}_{i\alpha}(\vec{\sigma}\times \vec{d^{0}_{ij}})^{z}_{\alpha\beta}c_{j\beta}+\ell\sum_{i\alpha}\mu_{i}E_{z}c^{\dag}_{i\alpha}c_{i\alpha},
\end{equation}
where $c^{\dag}_{i\alpha}$ ($c_{j\beta}$) creates (annihilates) an electron with spin polarization $\alpha$ ($\beta$) at site $i$ ($j$). The sum is taken over all pairs of nearest neighbor $\langle i,j\rangle$ or next-nearest-neighbor $\langle\langle i,j\rangle\rangle$ lattice sites. The first term in Eq. (1) is the usual nearest-neighbor hopping integral with a transfer energy $t=1.6$ eV. The second term describes the effective SO coupling with $\lambda_{SO}$=3.9 meV. $\vec{\sigma}=(\sigma_{x},\sigma_{y},\sigma_{z})$ is the vector of the Pauli matrices. $v_{ij}$ is determined by the orientation of the two nearest bonds connecting the next-nearest neighbors, i.e., $v_{ij}=+1$ ($-1$) if the next-nearest-neighbor hopping
is anticlockwise (clockwise) with respect to the positive z axis. The third term is the intrinsic
Rashba SO coupling with $\lambda_{R2}=0.7$ meV, where $\vec{d^{0}_{ij}}=\vec{d_{ij}}/|d_{ij}|$ with $\vec{d_{ij}}$ connects two sites $i$ and $j$ in the same sublattice, and $u_{ij}=\pm1$ represents the A(B) site. The fourth term is the staggered sublattice potential energy produced by the external electric field, where $\mu_{i}=\pm1$ for the A(B) site. The tight-binding Hamiltonian can also be used to describe other group-IV elements, such as germanium with a different set of parameters ($t=1.3$ eV, $\lambda_{SO}$=43 meV, $\lambda_{R2}$=10.7 meV, and $l$=0.33 ${\AA}$ in the case of Ge \cite{CChen:2012}).

When the system is perturbed by a time-dependent Coulomb potential, the electron-electron interactions would induce a charge density fluctuation which acts to screen the external perturbation. This dielectric screening determines the normal modes of the charge density oscillations. Electronic excitations are characterized by the transferred momentum $q$ and the excitation frequency $\omega$, which determine the dielectric function. Within the random-phase approximation, the dielectric function can be expressed as:
\begin{eqnarray}
\epsilon(q,\omega)=\epsilon_{0}-v_{q}\chi^{0}(q,\omega)
\end{eqnarray}
, where $v_{q}=2\pi e^{2}/q$ is the in-plane Fourier transformation of the bare Coulomb potential energy, and $\epsilon_{0}$=2.4 (taken from graphite \cite{Shung:1986}) is the background dielectric constant due to the deep-energy electronic states. It is noted that changing $\epsilon_{0}$ would make a vertical shift of the real part of dielectric function (Fig. 3) and thus change the peak intensity and position in the energy loss function (Fig. 4). However, the plasmon behavior is qualitatively the same (Fig. 5). Such a form of dielectric function is utilized in many researches on two-dimensional systems, theoretically \cite{Wu:2011, Yuan:2011, Pisarra:2014} and experimentally \cite{Shin:2011, Eberlein:2008, Kramberger:2008}. The bare response function $\chi^{0}$ is given by \cite{Shung:1986,Ehrenreich:1959}:

\begin{eqnarray}
\chi^{0}(q,\omega)&=&\sum_{n,n'}\int_{1stBZ}\frac{dk_{x}dk_{y}}{(2\pi)^{2}}|\langle n^{\prime}; \vec{k}+\vec{q}|e^{i\vec{q}\cdot \vec{r}}|n;\vec{k}\rangle|^{2}
\nonumber\\
&\times &
\frac{f(E^{n'}(\vec{k}+\vec{q}))-f(E^{n}(\vec{k}))}{E^{n'}(\vec{k}+\vec{q})
-E^{n}(\vec{k})-(\omega +i\Gamma)}\ .
\label{polarization}
\end{eqnarray}
Each Bloch state is labeled by the Bloch wave vector $k$ and the band index $n$. $E^{n}$ is the corresponding eigenvalue. The Fermi-Dirac distribution is $f(E^{n})=1/\{1+exp[(E^{n}-\mu)/k_{B}T]\}$, where $k_{B}$ is the Boltzmann constant and $\mu$ is the chemical potential. $\mu$ is set to zero over the range of the symmetric conduction ($c$) and valence ($v$) bands under the investigation. $\Gamma$ is the energy broadening which results from various de-excitation mechanisms, e.g., the optical transitions between the valence- and conduction-band states. It is usually set small in the low-frequency region. Changing $\Gamma$ would not alter the peak positions in the energy-loss function (Fig. 4), but the peak intensities and widths will be affected.

The details of the calculation of Coulomb matrix elements are shown below.
\begin{eqnarray}
|\langle n^{'};\vec{k}+\vec{q}|e^{i\vec{q}\centerdot\vec{r}}|n;\vec{k}\rangle=\nonumber\\
\sum_{s=\alpha,\beta}\sum_{i=1,2}\langle \phi_{z}(\vec{r}-\vec{\tau_{i}})|e^{-i\vec{q}\cdot(\vec{r}-\vec{\tau_{i}})}|\phi_{z}(\vec{r}-\vec{\tau_{i}})\rangle
[u_{n^{'}si}(\vec{k}+\vec{q})u_{nsi}^{*}(\vec{k})].
\end{eqnarray}
$\tau_{1}$ and $\tau_{2}$ define the positions of atoms in a unit cell. $u_{nsi}(\vec{k})$ ($u_{n^{'}si}(\vec{k}+\vec{q})$) are the coefficients for the TB wave functions derived from Eq. (1). $\langle\phi_{z}(\vec{r}-\vec{\tau_{i}})|e^{-i\vec{q}\cdot(\vec{r}-\vec{\tau_{i}})}|\phi_{z}(\vec{r}-\vec{\tau_{i}})\rangle=I(q)=[1+[\frac{qa_{0}}{Z}]^{2}]^{-3}$ was calculated by using the hydrogenic wave function \cite{Shung:1986}, where $a_{0}$ is the Bohr radius and $Z$ is an effective core charge \cite{Zener:1930}. For small $q^{,}s$, I(q) is very close to 1. The double integral is performed numerically with respect to wave vector and band index.

\bigskip
\bigskip
\centerline {\textbf {III. RESULTS AND DISCUSSION}}%
\bigskip
\bigskip

The Hamiltonian of Eq. (1) is solved numerically to derive the whole $\pi$-band structure and the coefficients for the TB wave functions. The low-energy band structure around the $K$ point in the 1st Brillouin zone is presented in Fig. 2. In this region, the energy dispersion can be described by Dirac theory \cite{Ezawa:2012}. At zero fields (Fig. 2(a)), all energy bands are spin degenerate. The main effect of the SOI is to produce a small energy gap $E_{g}$ between the valence and conduction bands. When a perpendicular electric field $E_{z}$ is applied, each band exhibits the spin-split subbands due to the breaking of the inversion symmetry (Figs. 2(b)-2(d)). Consequently, the system is characterized by two energy gaps, $E_{1g}$ and $E_{2g}$ for spin-down (blue curve) and spin-up (red curve) subbands, respectively. The former decreases with increasing field strength, while the opposite is true for the latter. The spin-down and spin-up subbands are reversed at different valleys. Whenever $E_{z}=E_{c}$ ($\approx\lambda_{SO}/l=17\ meV{\AA}^{-1}$), the staggered sublattice potential energy exactly compensates for the SOI on the spin-down states and the gap of $E_{1g}$ is closed (Fig. 2(c)). Meanwhile, the spin-split energy at the $K$ point reaches its maximum, i.e., $2\lambda_{SO}$. For $E_{z}>E_{c}$ (Fig. 2(d)), $E_{1g}$ is reopened, and the system is transformed into a band insulator. After that, both $E_{1g}$ and $E_{2g}$ increase with $E_{z}$, while the spin-split energy at $K$ point remains constant at $2\lambda_{SO}$.

The unscreened-excitation spectrum is helpful in understanding the single-particle excitation (SPE) channels. At zero temperature (Fig. 3(a)), the interband excitations are the only available excitation channel. The threshold excitation energy is expressed as $\omega^{inter}_{th}=2\sqrt{(\lambda_{SO})^{2}+v_{F}^{2}q^{2}/4}$.
The real ($\epsilon_{1}$) and imaginary ($\epsilon_{2}$) parts of the longitudinal dielectric function, related through the Kramers-Kronig relations, exhibit a logarithmic-divergent peak and a step-like structure at $\omega^{inter}_{th}$, respectively. The interband-gap transitions account for these prominent structures. There are no zero points in $\epsilon_{1}$. That is, for $T=0$, no self-sustained plasmon excitations can exist in the low-frequency region. The intraband transitions are allowed at finite temperatures; these transitions create additional special structures in $\epsilon$, i.e., an asymmetric dip and peak in $\epsilon_{1}$ and $\epsilon_{2}$, respectively, as illustrated in Fig. 3(b) for T=50 K. At the same time, the interband-transition-related structures become weaker. The highest intraband-excitation energy is $\omega^{intra}_{ex}=v_{F}q$, which is lower than $\omega^{inter}_{th}$. Therefore, a SPE gap is developed in $\epsilon_{2}$. The asymmetric dip in $\epsilon_{1}$ creates two zero points. One of the zero points is located in the $\epsilon_{2}$ gap (only for the condition $k_{B}T<E_{g}$), where an undamped plasmon mode can occur. The temperature range that allows the undamped plasmon depends on the width of the SPE gap; this width is tunable by an external E-field.

The effects of the E-field on $\epsilon$ are displayed in Figs. 3(c)-3(e) for fixed $T=50$ K and different $E_{z}$'s. At $E_{z}=0.5E_{c}$, shown in Fig. 3(c), the step structure (peak) in $\epsilon_{2}$ ($\epsilon_{1}$) is split into two owing to the lifting of the spin degeneracy, as indicated by the pair of blue arrows. The two steps of the structure develop at $\omega^{inter}_{1,th}$ and $\omega^{inter}_{2,th}$, respectively. Between $\omega^{inter}_{1,th}$ and $\omega^{inter}_{2,th}$ the density of the e-h pairs is half of that beyond $\omega^{inter}_{2,th}$. This means the creation of a low-damped plasmon region. $\omega^{inter}_{1,th}$ decreases when $E_{z}$ is increased from zero to $E_{c}$, i.e., the SPE gap is reduced. This may hinder the occurrence of an undamped plasmon. The opposite is true for $E_{z}>E_{c}$ (Fig. 3(e)). At $E_{z}=E_{c}$ (Fig. 3(d)), $\omega^{inter}_{1,th}$ and $\omega^{intra}_{ex}$ merge. Therefore, both zero points of $\epsilon_{1}$ are located where $\epsilon_{2}$ is finite. Under such condition, only damped plasmons can exist.

The energy-loss function ($\propto$ $Im[-1/\epsilon]$), related to the screened excitation spectrum, is used to understand the collective-excitation mode of the electrons. The quantity can be measured by the electron energy loss spectroscopy in transmission \cite{Eberlein:2008, Kramberger:2008,Wachsmuth:2013,Pan:2012} and reflection \cite{Langer:2009}, and inelastic light scattering \cite{Richards:2000}. Each prominent structure in $Im[-1/\epsilon]$ may be viewed as a plasmon excitation with different
degrees of Landau damping. At T=0, the low-frequency $Im[-1/\epsilon]$ appears as plateaus without and with an electric field, as shown in the insets of Figs. 4(a) and 4(b), respectively. The featureless plateau structures correspond to the interband excitations, which exhibit single-particle like behaviors. At finite temperatures, a prominent peak is induced with a frequency lower than the plateaus (Figs. 4(a)-4(d) for $T=50$ K). The corresponding dielectric functions reveal that this peak could arise from the interplay between the intraband and interband-gap transitions (Figs. 3(b), 3(c), and 3(e)) or merely from the sufficiently strong intraband transitions under the condition of a zero SPE gap (Fig. 3(d)). The former leads to the stronger peak intensity and the narrower peak width in the energy-loss function (Figs. 4(a), 4(b), and 4(d)), compared with that caused by the latter (Fig. 4(c)).

The $q$-dependent behavior of plasmons is important in understanding their characteristics. Figs. 5(a)-5(c) show the results for different $E_{z}'s$ at fixed $T=50$ K. The color scale stands for the values of the energy-loss function. The dashed curves in each figure represent the highest intraband (white curves) and the lowest interband (blue curves) SPE energies, which obey the relations $\omega=v_{F}q$ and $\omega= 2\sqrt{(\ell E_{z}-s\sqrt{\lambda_{SO}^{2}+a^{2}\lambda^{2}_{R}q^{2}/4})^{2}+\hbar^{2}v_{F}^{2}q^{2}/4}$
with $s=\pm 1$, respectively. The expressions are derived from the analytical solution of single-particle spectrum near the K point \cite{Ezawa:2012}. These boundary curves define the various regions with different degrees of Landau damping. Between the white and the blue boundary curves lies the SPE gap, where exists an undamped plasmon branch. The plasmon frequency $\omega_{p}$ is well fitted by $\sqrt{q}$ in the long-wavelength limit, a feature of a 2D plasmon. For lower temperatures (Figs. 5(a) and 5(b)), this plasmon mode is finally damped out in the intraband SPE continuum at larger $q's$. The existence range of the undamped plasmon in momentum is proportional to the width of the SPE gap, which is reduced in the region of $0<E_{z}<E_{c}$ (Fig. 5(b) for $E_{z}=0.5E_{c}$). There are two blue boundary curves in $E_{z}\neq 0$ (Figs. 5(b)-5(d)). They are the result of the lifting of the spin degeneracy. Between the two blue curves lies a low damped region. In Fig. 5(b), a broad plasmon branch appears in the region, just above the undamped plasmon mode. The coexistence of the two collective-excitation modes can only be realized at $k_{B}T\approx E_{1g}$. The SPE gap is closed at $E_{z}=E_{c}$ (Fig. 5(c)). Then, the plasmon is damped at arbitrary $q's$. Until $E_{z}>E_{c}$, the undamped plasmon branch is regained (not shown). Under the above conditions, the plasmon dispersion is confined between two SPE boundary curves. However, when the temperature is sufficiently high, e.g., $T=200$ K and $E_{z}=0.5E_{c}$ in Fig. 5(d), the plasmon branch crosses the SPE boundaries and shows abrupt changes in intensity and bandwidth. The discontinuous plasmon dispersion is more evident when the plasmon crosses the upper blue curve. After that, the plasmon enters the strong damped region. It is worth noting that the temperature-induced plasmon differs from the extrinsic plasmon discussed in doped silicene \cite{ChangHR:2014}. The dispersion relation of the extrinsic plasmon is decided by the Fermi energy. This is a great contrast to the intrinsic plasmon whose dispersion is dominated by the temperature. The different electron distributions result in the different plasmon dispersions between in the two conditions.

The effect of temperature variation on the plasmon excitation is worth a further discussion. Figs. 6(a)-6(c) show the T-dependent plasmon spectra at fixed $q=1$ and different $E_{z}'s$. The distribution range of the undamped plasmon mode (between the blue and white dashed curves) in $T$ is reduced in the region of $0<E_{z}<E_{c}$ (Fig. 6(b) for $E_{z}=0.5E_{c}$) and diminished to zero at $E_{z}=E_{c}$ (Fig. 6(c)), compared to the condition of $E_{z}=0$ (Fig. 6(a)). The undamped plasmon is transformed to a damped one above a critical temperature ($T_{c}\simeq 100$ K for $E_{z}=0$). The transformation in the plasmon mode during the change in $T$ is a discontinuous process, a feature that could allow for the identification of the spin-orbit energy gap. The discontinuous dispersions of the plasmons remain when changing background dielectric constant $\epsilon_{0}$ and energy broadening $\Gamma$. In Fig. 6(b) for $E_{z}=0.5E_{c}$, two critical temperatures exist through which the plasmon spectrum is drastically altered. The two $T_{c}'s$ correspond to the two energy gaps opening in the system. This could be used to examine the spin degeneracy and determine $E_{c}$. Note that above the highest $T_{c}$ ($\approx175$ K for $E_{z}=0.5E_{c}$), the plasmon intensity grows monotonically with $T$, similar to the behavior of the intrinsic plasmon in gapless graphene \cite{Lin:2000}. This indicates that at such high temperatures the SOI effect becomes negligible.

For fixed $q$ and $T$, the plasmon existence and its spectral weight can be effectively controlled by varying $E_{z}$, as shown in Fig. 7. At $T=50$ K and $q=1$ (Fig. 7(a)), the undamped plasmon is stable for $E_{z}<0.5E_{c}$ due to the significantly large SPE gap. When $E_{z}$ is increased to near $E_{c}$, the plasmon is damped by the interband SPEs and faints. The SPE gap is reopened in the region of $E_{z}>E_{c}$. After that the plasmon is intensified first with the increment of $E_{z}$, and when $E_{z}$ exceeds another critical value $E^{'}_{c}$ ($E^{'}_{c}\approx3E_{c}$ for $T=50$ K and $q=1$), the plasmon is gradually damped out in the intraband SPE continuum. At $T=100$ K (Fig. 7(b)), the undamped plasmon barely exists in $E_{z}<2E_{c}$, because the overly strong intraband transitions make the plasmon frequency higher than the threshold interband SPE energy. Until $E_{z}>2E_{c}$, the energy gap becomes larger than $k_{B}T$, and the undamped plasmon emerges. The undamped plasmon at $T=100$ K can survive in a larger $E_{z}$ than that at $T=50$ K. The stronger intraband transitions can explain this. At $T=200$ K (Fig. 7(c)), the plasmon spectrum displays three distinct degrees of Landau damping in different $E_{z}$ ranges. In the first region ($0<E_{z}<0.7E_{c}$), the plasmon is located beyond the two interband SPE boundaries and suffers strong Landau damping. In the second region ($0.7E_{c}<E_{z}<2.5E_{c}$), the plasmon lies between the two interband boundaries and is weakly damped. In the third region ($E_{z}>2.5E_{c}$), the plasmon lies inside the SPE gap and is undamped. Though the plasmon intensity strongly depends on $E_{z}$, the change in the plasmon frequency is insignificant. This illustrates that the appearance of the low-frequency plasmon is mainly dominated by the intraband transitions.

\bigskip
\bigskip
\centerline {\textbf {IV. SUMMARY AND CONCLUSIONS}}%
\bigskip
\bigskip

The Coulomb-excitation properties of silicene are studied by the tight-binding model and random-phase approximation. The temperature and E-field effects are considered simultaneously. At finite temperatures, an undamped plasmon is induced by the interplay between the intraband and interband-gap transitions. The plasmon mode only exists in a limited region of $T$ and $q$, which can be further tuned by varying $E_{z}$. Thanks to the opening of two energy gaps from the spin split in $E_{z}$, there are three distinct degrees of Landau damping in different $q$, $T$, and $E_{z}$ ranges. Therefore, the plasmon dispersion is discontinuous at certain critical $q's$ and $T's$. The phenomenon could be used to identify the spin degeneracy and to determine relevant theoretical parameters, such as the magnitude of SOI. We further show that, for given $q$ and $T$, changing $E_{z}$ can control the plasmon intensity and its existence at a nearly fixed frequency, a property that should be important in plasmonic applications. The tunable low-frequency plasmons can potentially be also found in other 2D crystals with band gaps that can be adjusted by an external E-field, e.g., germanene and stanene.

\bigskip

\bigskip

\centerline {\textbf {ACKNOWLEDGMENT}}%

\bigskip

\bigskip

\noindent \textit{Acknowledgments.} This work was supported by the NSC of Taiwan, under Grant No. NSC 102-2112-M-006-007-MY3.

\newpage

\par\noindent ~~~~$^\star$e-mail address: yarst5@gmail.com

\par\noindent ~~~~$^\dag$e-mail address: ggumbs@hunter.cuny.edu

\par\noindent ~~~~$^\dag$$^\dag$e-mail address: mflin@mail.ncku.edu.tw

\bigskip \vskip0.6 truecm

\noindent

\newpage

\centerline {\Large \textbf {Figure Captions}}

\vskip0.3 truecm

Figure 1. The lattice geometry of low-buckled silicene. The two sublattices are separated by a perpendicular distance $2l$. The staggered sublattice potential energy is produced by an external electric field $E_{z}$.

Figure 2. The low-energy bands of monolayer silicene in the presence of spin-orbit coupling about the K point are shown in (a). Those under $E_{z}=0.5E_{c}$, $E_{c}$, and $2E_{c}$ are shown in (b), (c), and (d), respectively.

Figure 3. The dielectric function as a function of $\omega$ at $q=1$ and $T=0$ is shown in (a). Those for $T=50$ K and different $E_{z}'s$ are shown in (b)-(e). The purple and orange curves represent the real and the imaginary parts, respectively. The energy broadening width is $\Gamma=0.1$ meV. $q$ is in unit of $10^{5}/cm$.

\bigskip

Figure 4. The energy-loss functions for fixed $q=1$, $T=50$ K, and different $E_{z}'s$ are plotted in (a)-(d). Those for $E_{z}=0$ and $E_{z}=0.5E_{c}$ at $T=0$ are plotted in the insets of (a) and (b), respectively.

\bigskip

Figure 5. The q-dependent plasmon spectrum at $T=50$ K for (a) $E_{z}=0$, (b) $E_{z}=0.5E_{c}$, (c) $E_{z}=E_{c}$, and (d) $E_{z}=2E_{c}$. The dashed blue and white curves in each panel indicate the interband and intraband SPE boundaries, respectively. The color scale stands for the values of the energy-loss function.

\bigskip

Figure 6. The T-dependent plasmon spectrum at fixed $q=1$ and different $E_{z}'s$: (a) $E_{z}=0$, (b) $E_{z}=0.5E_{c}$, and (c) $E_{z}=E_{c}$. The blue and white dashed curves in each panel indicate the interband and the intraband SPE boundaries, respectively. The color scale stands for the values of the energy-loss function.

Figure 7. The $E_{z}$-dependent plasmon spectrum at fixed $q=1$ and different $T's$: (a) $T=50$ K, (b) $T=100$ K, and (c) $T=200$ K. The color scale stands for the values of the energy-loss function.

\end{document}